\documentclass[prl,twocolumn,showpacs]{revtex4}
\usepackage{graphicx}
\usepackage{graphics}
\usepackage{amsfonts}
\usepackage{amsmath}
\usepackage{epsfig}

\begin{document}
\catcode`\ä = \active \catcode`\ö = \active \catcode`\ü = \active
\catcode`\Ä = \active \catcode`\Ö = \active \catcode`\Ü = \active
\catcode`\ß = \active \catcode`\é = \active \catcode`\è = \active
\catcode`\ë = \active \catcode`\ô = \active \catcode`\ê = \active
\catcode`\ø = \active \catcode`\ò = \active \catcode`\í = \active
\catcode`\Ó = \active \catcode`\ú = \active \catcode`\á = \active
\catcode`\ã = \active
\defä{\"a} \defö{\"o} \defü{\"u} \defÄ{\"A} \defÖ{\"O} \defÜ{\"U} \defß{\ss} \defé{\'{e}}
\defè{\`{e}} \defë{\"{e}} \defô{\^{o}} \defê{\^{e}} \defø{\o} \defò{\`{o}} \defí{\'{i}}
\defÓ{\'{O}} \defú{\'{u}} \defá{\'{a}} \defã{\~{a}}



\newcommand{\li}{$^6$Li}
\newcommand{\na}{$^{23}$Na}
\newcommand{\vect}[1]{\mathbf #1}
\newcommand{\g}{g^{(2)}}
\newcommand{\one}{|1\rangle}
\newcommand{\two}{|2\rangle}
\newcommand{\V}{V_{12}}
\newcommand{\kfa}{\frac{1}{k_F a}}

\title{Condensation of Pairs of Fermionic Atoms Near a Feshbach Resonance}

\author{M.W. Zwierlein, C.A. Stan, C.H. Schunck, S.M.F. Raupach, A.J. Kerman, and W. Ketterle}

\affiliation{Department of Physics\mbox{,} MIT-Harvard Center for
Ultracold Atoms\mbox{,}
and Research Laboratory of Electronics,\\
MIT, Cambridge, MA 02139}

\date{\today}

\begin{abstract}
We have observed Bose-Einstein condensation of pairs of fermionic
atoms in an ultracold $^6$Li gas at magnetic fields above a
Feshbach resonance, where no stable $^6$Li$_2$ molecules would
exist in vacuum. We accurately determined the position of the
resonance to be 822$\pm$3 G. Molecular Bose-Einstein condensates
were detected after a fast magnetic field ramp, which transferred
pairs of atoms at close distances into bound molecules. Condensate
fractions as high as 80\% were obtained. The large condensate
fractions are interpreted in terms of pre-existing molecules which
are quasi-stable even above the two-body Feshbach resonance due to
the presence of the degenerate Fermi gas.
\end{abstract}
\pacs{03.75.Ss,05.30.Fk}

\maketitle

Ultracold atomic gases have become a medium to realize novel
phenomena in condensed matter physics and test many-body theories
in new regimes.  The particle densities are $10^8$ times lower
than in solids, but at temperatures in the nanokelvin range, which
are now routinely achieved, interactions and correlations become
important. Of particular interest are pairing phenomena in
fermionic gases, which have direct analogies to
superconductivity~\cite{stoo99var}.

The interactions which drive the pairing in these gases can be
controlled using a Feshbach resonance \cite{feshbach}, in which a
molecular level is Zeeman-tuned through zero binding energy using
an external magnetic field. This provides an opportunity to
experimentally probe what is known as the BCS-BEC crossover; as
the strength of the effective attractive interaction between
particles is increased a continuous transition from condensation
of delocalized Cooper pairs to tightly-bound bosonic molecules is
predicted \cite{BECBCS,atomBECBCS,ohas02,falc04}. Whereas in the
BCS limit the pairing is a strictly many-body effect
\cite{houb97}, in the BEC limit a pair of fermions is bound even
as an isolated molecule. A novel form of high-temperature
superfluidity has been predicted to emerge in the crossover region
\cite{BECBCS,atomBECBCS,ohas02,falc04}. Until recently, the
observation of condensation phenomena in fermionic atomic gases
was restricted to the extreme BEC limit, where several groups have
observed Bose-Einstein condensation of diatomic molecules
\cite{joch03bart04, grei03, zwie03molBEC,bour04}.

An important step was recently reported, in which condensation of
atomic $^{40}$K fermion pairs was observed on the BCS side of a
Feshbach resonance \cite{rega04}. It was argued that those pairs
were not bound into molecules, but merely moved together in a
correlated fashion, similar to Cooper pairs of electrons in a
superconductor \cite{press}. However, the exact nature of these
pairs remained unclear. In this paper, we apply similar techniques
to \li\ atoms, which have very different collisional properties
\cite{houb98}, and observe the pair condensation phenomenon above
a Feshbach resonance. In contrast to the previous work, where at
most 15\% of the atom pairs were condensed \cite{rega04},
condensate fractions of up to 80\% were observed. We argue that
such a high condensate fraction is unlikely for pairs which are
long-range, but rather indicates a condensate of short-range atom
pairs which are essentially molecular in character even on the BCS
side of the resonance.

A simple argument supports this possibility. In the basic picture
of a Feshbach resonance, a molecular state above the dissociation
threshold has a finite lifetime, which becomes shorter as the
energy of the state increases, as recently observed \cite{muka04}.
In the presence of the Fermi sea, its lifetime will be increased
due to Pauli blocking. The molecular level will be populated until
its energy becomes larger than twice the Fermi energy
corresponding to the total number of atoms. The BEC-BCS crossover
is expected to occur at this point, and not at the location of the
two-body Feshbach resonance~\cite{ohas02,falc04}.

The basic setup of our experiment was described in
\cite{zwie03molBEC}. By sympathetic cooling of \li{ } atoms with
$^{23}$Na in a magnetic trap, a degenerate gas of about
$3\times10^7$ \li\ fermions at $\sim$0.3 $T/T_F$ was created.
After transfer into an optical dipole trap (maximum power 9 W
focused to an $e^{-2}$ radius of 25 $\mu$m),  an equal mixture of
atoms in the lowest two hyperfine states $\one$ and $\two$ was
prepared. The sample was evaporatively cooled at a magnetic field
in the range from 770-810 G using an exponential ramp down
(timescale $\sim$400 ms) of the optical trap to a final laser
power of 15 mW. This created essentially pure Bose-Einstein
condensates of up to $3\times10^6$ $^6$Li$_2$ molecules. The
observed trap vibrational frequencies could be described by the
following expression: $\nu_{rad}\approx$ 115 Hz $\sqrt{P}$,
$\nu_{ax}\approx$ 1.1 Hz $\sqrt{P+120B}$ where $P$ is the optical
power in mW, and $B$ is the magnetic field in kG. The latter
dependence arises from the residual axial curvature of the
magnetic field. Considerable improvements over our previous setup
\cite{zwie03molBEC} led to an improved e$^{-1}$ condensate
lifetime of 10 s at 770 G. Moreover, within the experimental
uncertainty in the total number of molecules ($\sim$50\%), mean
field measurements were consistent with a molecule-molecule
scattering length of $0.6 a$ where $a$ is the atomic scattering
length \cite{petr03dimers,newtrap}.

\begin{figure}
    \begin{center}
    \includegraphics[width=3.5in]{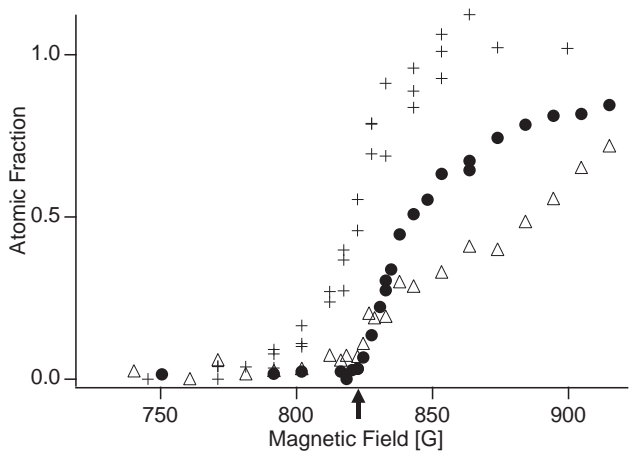}
    \caption[Title]{Determination of the Feshbach
    resonance position.  Shown is the onset of molecule dissociation when the
    magnetic field was slowly raised, and then ramped down to zero field with a variable
    rate: Using a switch-off of the power supply at an initial rate of 30 G/$\mu$s (crosses), a linear ramp to zero field of 100 G/ms (circles), a linear ramp of 16 ms at 12.5
    G/ms, followed by switch-off (triangles). The identical threshold for the two lowest ramp
    rates determines the resonance position to be 822$\pm$3 G, marked by an arrow.} \label{fig:resonance}
    \end{center}
\end{figure}

Previously, the location of the \li{ } Feshbach resonance was
determined either by observing a peak in the inelastic loss
\cite{diec02fesh}, or the interaction energy of a
$|1\rangle-|2\rangle$ mixture \cite{bour03}. A more accurate
determination can be made by mapping out the onset of dissociation
of the molecular state \cite{muka04,rega04}. After releasing an
almost pure molecular sample from the trap at 770 G, the magnetic
field was linearly ramped up in 10 ms to a variable value. During
that time, the particle density dropped by a factor of 1000. If
the field crossed the resonance, molecules dissociated into atoms.
These atoms were then imaged at zero field, where the remaining
molecules were not detected \cite{zwie03molBEC}. The Feshbach
resonance appeared as a sharp onset in the number of detected
atoms (Fig.~\ref{fig:resonance}). The speed of the downward ramp
to zero field had to be chosen carefully. Fast ramps could
dissociate very weakly-bound molecules \cite{cubi03}, such that
the Feshbach resonance appeared systematically shifted to lower
fields. For too slow a ramp down, on the other hand, we found that
even for clouds as dilute as $\sim3\times10^{10}$ cm$^{-3}$
molecules were recreated, lowering the measured atomic fraction.
However, when we varied the ramp rate over more than three orders
of magnitude, we found a range of rates which gave identical
thresholds at 822$\pm$3 G (Fig.~\ref{fig:resonance}).

To produce samples in the crossover region, we started with an
essentially pure Bose-Einstein condensate of molecules formed at
790 G. The laser power of the optical trap was increased in 500 ms
from 15 mW to 25 mW in order to accommodate larger Fermi clouds
above the resonance. In some experiments, we used a deeper trap
with up to 150 mW of power; the additional compression was carried
out after ramping in 500 ms to 900 G to avoid enhanced losses on
the BEC side of the resonance. Once the final trap depth was
reached, the magnetic field was ramped in 500 ms to values from
650 to 1025 G. The adiabaticity of this ramp was checked by
ramping back to 790 G and observing identical density profile and
condensate fraction, similar to studies in Ref.
\cite{joch03bart04}. At 1025 G, the total peak density of the spin
mixture in the deepest trap was $3 \times10^{13}\rm\,cm^{-3}$,
corresponding to a Fermi energy of 3.6 $\mu$K and inverse
Fermi-wavevector $k_F^{-1} \simeq 2000\,a_0$, where $a_0$ denotes
the Bohr radius.

To probe the gas, we released it from the trap, and after a
variable delay $\tau_d$ of usually 40 $\mu$s, applied a rapid
transfer technique \cite{rega04}: the magnetic field was switched
off exponentially to zero with an initial slew rate of 30
G/$\mu$s, which adiabatically converted pairs of atoms into deeply
bound molecules at zero field~\cite{fastramp}. As long as no
collisions or other dynamics occur during this ramp, the velocity
distribution of the resulting molecules then constituted a probe
of the atom pairs' center-of-mass motion before the measurement.
After 3-6 ms time-of-flight at zero field, we dissociated the
molecules with a 3 ms field pulse to 900 G and imaged the
resulting atoms after 2 ms at zero field
\cite{zwie03molBEC,reconv}. We could also selectively detect any
remaining atoms by omitting the dissociation pulse, and we
observed that for $\tau_d\le500$ $\mu$s, less than 10\% of the
sample consisted of atoms, independent of magnetic field. At
longer delay times, the atom-molecule conversion became less
efficient due to the decreased density.

\begin{figure}
    \begin{center}
    \includegraphics[width=3.5in]{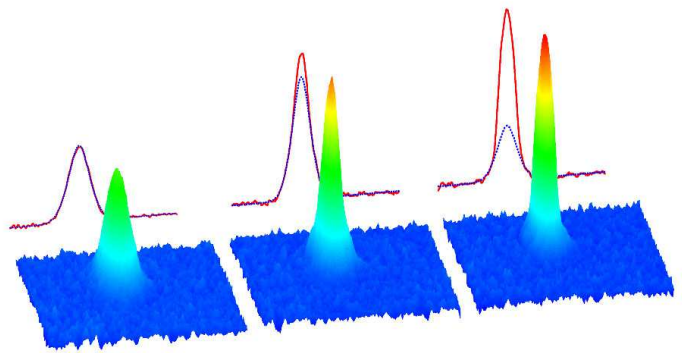}
    \caption[Title]{Emergence of a Bose-Einstein condensate of atom
    pairs as the temperature was lowered.  Shown are column densities (after 6 ms of time of flight) of the fermion mixture after a rapid transfer ramp from 900 G for
    three different initial temperatures  $T/T_F \approx 0.2$, 0.1 and 0.05, together with their
    axially integrated radial density profiles.  The dashed line is a
    gaussian fit to the thermal component. Condensate fractions are: 0.0, 0.1, and 0.6. Each cloud consists of about 2 million molecules. The
    field of view is 3 mm $\times$ 3 mm.} \label{fig:examplefits}
    \end{center}
\end{figure}

\begin{figure}
    \begin{center}
    \includegraphics[width=3.5in]{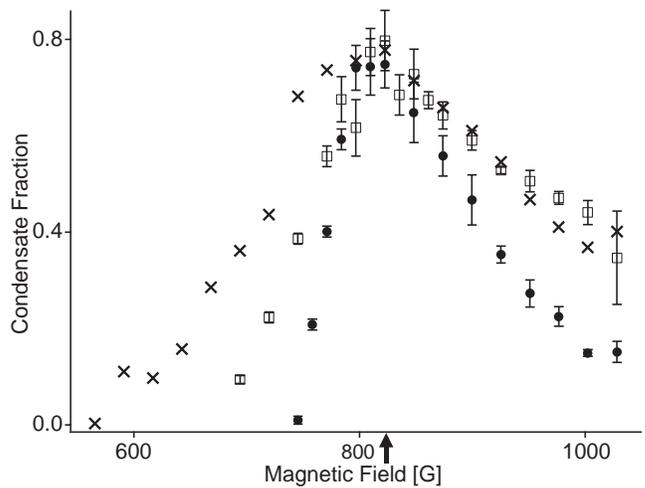}
    \caption[Title]{Condensate fraction after the rapid transfer vs. initial magnetic
    field, for different hold times at that field in the shallow trap ($P=25$ mW). Crosses: 2 ms
    hold time, after 500ms ramp to 1000 G and 4ms ramp to the
    desired field; squares and circles: 100 ms and 10 s hold time, after
    500 ms ramp from 790 G. The reduction of the condensate fraction for long
hold times far on the left side of the resonance is probably due to
the rapidly increasing inelastic losses for the more tightly bound
molecules \cite{cubi03,joch03lith}.  The lower condensate fraction at
high field for long hold times is probably an effect of lower density
since the number of atoms had decayed by a factor of 4 without change
in temperature.}
    \label{fig:CondFracholdtime}
    \end{center}
\end{figure}

\begin{figure}
    \begin{center}
    \includegraphics[width=3.5in]{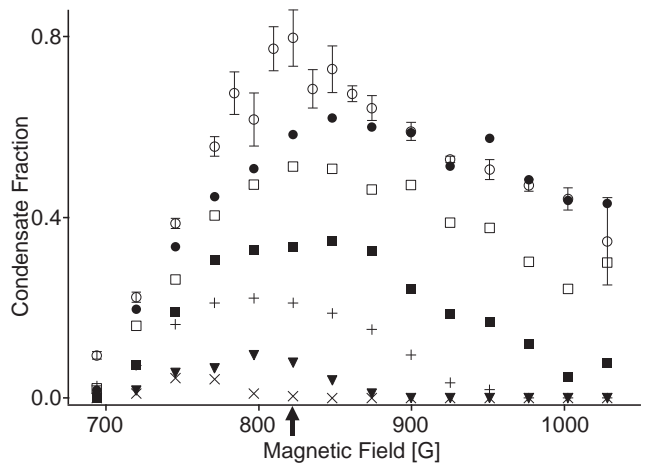}
    \caption[Title]{Condensate fraction for different temperatures as
    a function of magnetic field.  The temperature of the molecular
    cloud was varied by stopping the evaporative cooling
    earlier and applying parametric heating before ramping to the
    final magnetic field. Temperatures are parameterized by the
    molecular condensate fraction $N_0/N$ at 820 G (open circles: 0.8; filled circles: 0.58; open squares: 0.51; filled squares: 0.34; ``+": 0.21; triangles: 0.08; ``$\times$": $<$ 0.01).
    The lowest temperature was realized in the shallow trap ($P=25$ mW); the higher temperatures required a deeper trap ($P=150$ mW).}
    \label{fig:CondFractemperature}
    \end{center}
\end{figure}

\begin{figure}
    \begin{center}
    \includegraphics[width=3.5in]{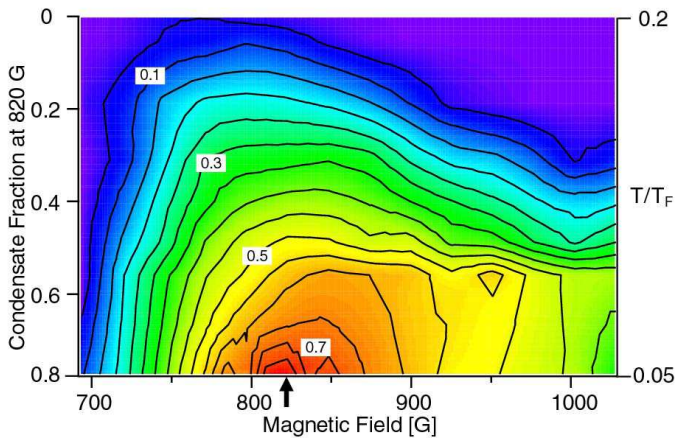}
    \caption[Title]{Temperature and magnetic field ranges over which pair condensation was observed (using the same data as in Fig.~\ref{fig:CondFractemperature}). The right
    axis shows the range in $T/T_F$ (measured at 1025 G)
    which was covered. For high degeneracies, fitting $T/T_F$ was
    less reliable and we regard the condensate fraction as a superior
    ``thermometer".  Note that for an isentropic cross-over from BEC to a
    Fermi sea, $T/T_F$ is approximately linearly related to the condensate fraction on the BEC
    side \cite{carr03}. For our maximum densities the region where $k_F|a|\ge1$ extends from about 710 G onward.
    }
    \label{fig:CondFracSurfaceplot}
    \end{center}
\end{figure}

Typical absorption pictures of molecular clouds after the rapid
transfer ramp are shown in Fig.~\ref{fig:examplefits} for
different temperatures, clearly exhibiting a bimodal distribution.
This is evidence for condensation of pairs of \li\ atoms on the
BCS side of the Feshbach resonance. The condensate fractions were
extracted from images like these, using a gaussian fit function
for the ``thermal" part and a Thomas-Fermi profile for the
``condensate". Figs.~\ref{fig:CondFracholdtime},
\ref{fig:CondFractemperature}, and \ref{fig:CondFracSurfaceplot}
show the observed condensate fraction as a function of both
magnetic field and temperature. The striking features of these
data are the high condensate fraction of 80\% near resonance, and
the persistence of large condensate fractions on the BCS side of
the resonance all the way to our maximum field of 1025 G. After 10
s hold time, this value was still as high as 20\%. These
observations were independent of whether the final magnetic field
was approached starting with a Fermi sea or a molecular
condensate. Note that for our peak densities, the strongly
interacting region of $k_F\left|a\right| > 1$ extends from 710 G
onward.

There is experimental evidence that the observed pair condensates
existed before the sweep, and were not produced during the sweep
by collisions. First, the observed condensate fraction depended on
the initial magnetic field. Second, the condensate fraction did
not change when we varied the delay time $\tau_d$ (between release
of the atoms from the trap and the magnetic field ramp) from 0
$\mu$s to 200 $\mu$s, although the density of the cloud changed by
a factor of $\sim$4~\cite{broadcond}. However, we cannot rule out
with certainty that the momentum distribution of the pairs is
modified by collisions during the ramp~\cite{ramp}. At our highest
densities, it takes about 4 $\mu$s to take the molecules created
during the ramp out of the strongly interacting region
($k_F|a|\ge1$). A classical gas at the Fermi temperature would
have a unitarity limited collision time comparable to the inverse
of the Fermi energy divided by $\hbar$, which is about 2 $\mu$s.
Furthermore, there may be Pauli blocking for the atoms, and
bosonic stimulation for the molecules.

Assuming that collisions during the ramp can be neglected, it is
still crucial to ask what exactly happens during the rapid
transfer ramp, and what kind of pairs would likely be detected.  A
reasonable assumption is that atoms form molecules preferentially
with their nearest neighbor, independent of the center-of-mass
velocity of the pair. If, as our data shows, a large fraction of
the detected molecules are in a zero-momentum state after the fast
transfer, this means that nearest neighbors had opposite momenta.
If the distance between the fermions with opposite momenta making
up each pair were comparable to or larger than the interatomic
distance (as in long-range Cooper pairs) one would not expect to
find high condensate fractions; on the contrary, the transfer into
a tightly bound molecular state would randomly pick one of the
nearest neighbors, resulting in a thermal molecule. We regard our
observed high condensate fractions as evidence for the existence
of condensed atomic pairs above the Feshbach resonance, which are
smaller in size than the interatomic distance, and therefore
molecular in character. Their stability may be affected by Pauli
blocking and mean field effects, but their binding should be a
two-body effect and not a many-body effect as in the case of
Cooper pairs.

In conclusion, we have observed $^6$Li$_2$ molecular Bose-Einstein
condensates after a fast downward magnetic field ramp starting
with equilibrium samples at fields on either side of the broad
\li\ Feshbach resonance. Since there are no truly bound molecular
states above the resonance, we tentatively interpret our results
as a Bose-Einstein condensate of pairs of atoms which are
molecular in character and stabilized by the existence of the
Fermi sea. This condensate would drain particles from the Fermi
sea and lead to a reduced atomic Fermi energy roughly equal to
half the energy of the molecular level \cite{ohas02,falc04}.
Indeed, both in Ref. \cite{joch03bart04} and in the present work,
a reduction in the size of the cloud was observed as the Feshbach
resonance was approached from above, which may be due to this
effect. In agreement with theoretical predictions \cite{falc04} we
have observed pair condensation in the regime where $T/T_F < 0.1 -
0.2$ and $k_F \left|a\right| > 1$
(Fig.~\ref{fig:CondFracSurfaceplot}). The exact nature of the atom
pairs remains to be elucidated; they could be related to virtual
states or scattering resonances in the continuum; they may turn
out to be the tight-binding limit of Cooper pairs. It is also
possible that the pair condensate is a superposition state of
molecules and Cooper pairs \cite{ohas02,falc04,atomBECBCS}. We
regard the characterizing feature of the BEC-BCS cross-over a
qualitative change of the pairing phenomenon which has not yet
been observed.

This work was supported by the NSF, ONR, ARO, and NASA. We would like
to thank Walter Hofstetter, Michele Saba and Zoran Hadzibabic for
stimulating discussions. S.\ Raupach is grateful to the Dr. J\"urgen
Ulderup foundation for financial support.


\begin{thebibliography}{}
\bibitem{newtrap} The smaller value reported in our previous work \cite{zwie03molBEC} was based on an incorrect extrapolation of trap frequencies to one hundred times lower power. The radial trap frequency was strongly influenced by multiple intensity maxima in the optical focus.
\bibitem{fastramp} According to Fig.~\ref{fig:resonance}, this fast ramp can dissociate weakly bound molecules at low density, but at high density we observed an atom-molecule conversion $\ge 90$\%.
\bibitem{reconv} At this point the cloud was sufficiently dilute that the final switch off of the field did not reconvert atoms into molecules.
\bibitem{ramp} In the $^{40}$K experiment, the ratio of the effective sweep rate through resonance and Fermi energy was similar \cite{rega04}.
\bibitem{broadcond} The condensate peak broadened for longer $\tau_d$ and could not be reliably discerned after 200 $\mu$s.
\end{thebibliography}

\end{document}